\documentclass[10pt,twocolumn,prl]{revtex4}

\usepackage{amsmath,amssymb}

\usepackage{bm}
\usepackage{ifpdf}

\usepackage{graphicx}

\def\ket#1{\left|#1\right\rangle}
\def\bra#1{\left\langle#1\right|}

\newcommand{\proj}[1]{\ket{#1}\bra{#1}}

\DeclareMathOperator{\tr}{tr} \DeclareMathOperator{\Sym}{Sym}

\newcommand{\Mod}[1]{\left|#1\right|}

\newcommand{\norm}[1]{\left|\left|#1\right|\right|}

\def\H{\mathcal{H}}

\def\S{\mathcal{S}}

\def\C{\mathcal{C}}

\makeatletter

\renewcommand{\paragraph}{\@startsection
        {section}
        {1}
        {\parindent}
        {0pt}
        {0pt}
        {\textit}}

\renewcommand{\section}[1]{\paragraph{#1}.---}

\makeatother

\begin{document}
\hspace{0cm}
\title{A Finite de Finetti Theorem for Infinite-Dimensional Systems}

\author{Christian D'Cruz}
\email{C.H.D-Cruz@rhul.ac.uk}
\author{Tobias J. Osborne}

\author{R\"{u}diger Schack}

\affiliation{Department of Mathematics, Royal Holloway, University
of London, UK}
\date{\today}
\begin{abstract}

We formulate and prove a de Finetti representation theorem for
finitely exchangeable states of a quantum system consisting of $k$
infinite-dimensional subsystems. The theorem is valid for states
that can be written as the partial trace of a pure state
$\proj{\Psi}$ chosen from a family of subsets $\{\C_n\}$ of the full
symmetric subspace for $n$ subsystems. We show that such states
become arbitrarily close to mixtures of pure power states as $n$
increases. We give a second equivalent characterization of the
family $\{\C_n\}$.
\end{abstract}

\maketitle

The classical de Finetti theorem \cite{dFi37, Dia80} is a
representation theorem for exchangeable probability distributions.
It is of fundamental importance for the analysis of repeated trials
in Bayesian statistics \cite{Savage1972}.  For positive integers $n$
and $k$, a joint probability distribution for $k$ random variables
is said to be {\em $n$-exchangeable}, or simply {\em finitely
exchangeable}, if it can be written as the marginal of a symmetric
distribution for $n$ variables. A distribution is said to be {\em
infinitely exchangeable} if it is $n$-exchangeable for all $n$. The
content of the de Finetti theorem is that any infinitely
exchangeable probability distribution can be written as a convex
mixture of power distributions \cite{dFi37,endnote}. Additionally,
finitely exchangeable distributions can be approximated by such
mixtures \cite{Dia80}.

\begin{figure}[b]
\includegraphics[width=2.317in]{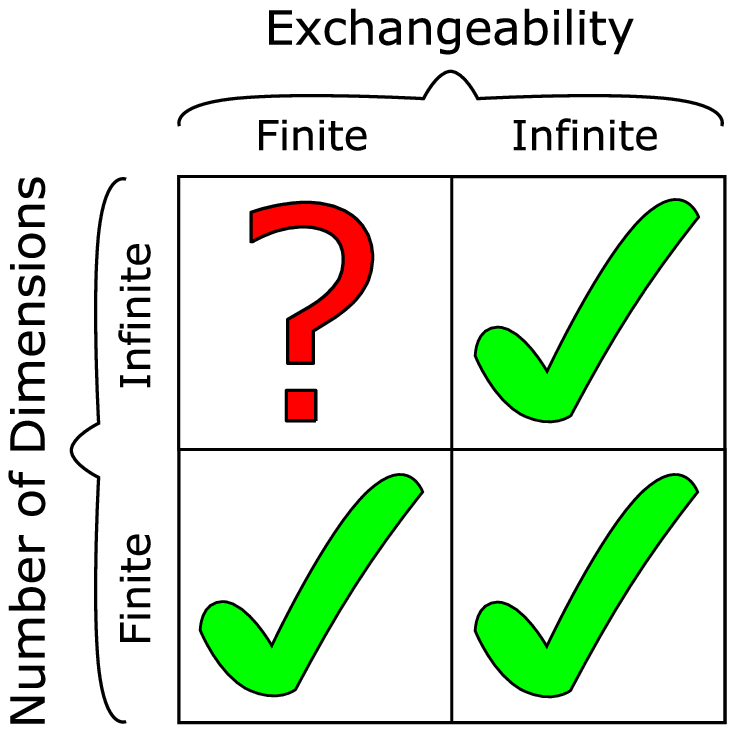}
\caption{Quantum generalizations of the classical de Finetti theorem
  fall into four classes. One can assume finite or infinite exchangeability,
  and the subsystems can be finite or \mbox{infinite-dimensional}.
  In the case of infinite exchangeability, we
  study a state $\rho_k$ that can be written as $\rho_k = \tr_{n - k} (\rho_n)$,
  for every integer $n>k$ for a symmetric state $\rho_n$.
  The theorems then state that $\rho_k$ can be written as a
  mixture of power states \cite{endnote24,HMo76,CFS02}. If finite exchangeability is assumed, the quantum de Finetti theorem says that a state
  $\rho_k =\tr_{n - k}(\rho_n)$, for a fixed value of $n$ and symmetric state
  $\rho_n$, can be approximated by a mixture of power states.
  The remaining case of finite exchangeability and infinite-dimensional subsystems is the topic of this Letter.
}\label{fig:progress}
\end{figure}
\begin{figure}[ht]

\includegraphics[width=3in]{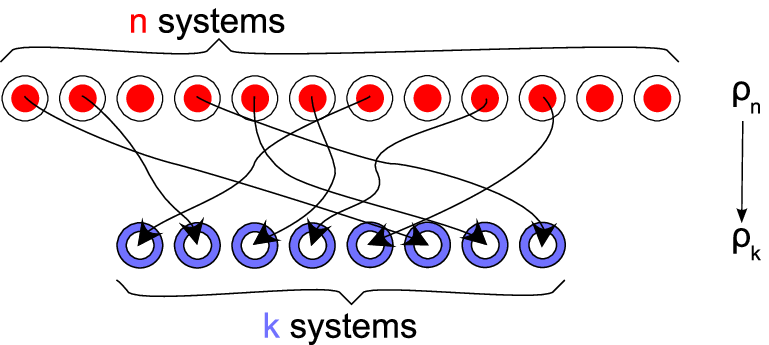}
\caption{A constructive illustration of symmetric states.  A state
\(\rho_n\) is symmetric when \(\pi \rho_n \pi^\dag = \rho_n\) for
all permutations \(\pi\).  In the figure, we illustrate these states
as those for which the state \(\rho_k\) is independent of the choice
of \(n-k\) subsystems to trace out, and of the order in which we
place the remaining \(k\).}\label{fig:contract}
\end{figure}

In recent years there has been increased interest in quantum
analogues of the de Finetti theorem.  Figure~\ref{fig:progress}
gives an overview of the possibilities.  They are of fundamental
interest in mathematics \cite{Stormer, HMo76, Petz}, quantum
information theory \cite{CFS02, Koe05,BrunEtAl, CHR06}, and quantum
foundations \cite{CFS02}. Concrete applications include quantum
state tomography \cite{CFS02, Bayes}, quantum process tomography
\cite{ScudoEtAl}, entanglement purification \cite{BrunEtAl}, and
quantum cryptography \cite{Ren05, Gis05}. Despite the progress in
this field, it remains an open question what quantum de Finetti
theorems exist for finitely exchangeable states on an array of
infinite-dimensional subsystems. A direct generalization of the
classical theorem to all finitely exchangeable quantum states is
impossible due to a counterexample given in \cite{CHR06}. There the
authors construct, for any integer $n>2$, an $n$-exchangeable state
on two infinite-dimensional subsystems that has a trace distance of
at least 1/2 from any mixture of power states \cite{endnote24}.

In this Letter we prove a quantum de Finetti theorem for a
particular class of $n$-exchangeable quantum states on a Hilbert
space $\H^{\otimes k}$, where $\mathcal{H}$ is infinite-dimensional,
and where $n$ and $k$ are arbitrary. Our class consists of all those
pure states $\rho_n = \ket{\psi}\bra{\psi}$ where $\ket{\psi}$ can
be written as a superposition of the form $\int d\gamma \,c_\gamma
\ket{\gamma}^{\otimes n}$ and each $\ket{\gamma}$ is a coherent
state.  It should be noted that such a superposition is very
different from a mixture of power states. This class contains many
physically relevant states, the simplest example being the
Schr{\"o}dinger cat states $|\alpha\rangle^{\otimes n} +
|\beta\rangle^{\otimes n}$ of $n$ optical modes. While such states
are covered by the finite de Finetti theorem, in reality they are
often an approximation to a continuous superposition $\int d\gamma\,
c_\gamma |\gamma\rangle^{\otimes n}$ where $c_\gamma$ is strongly
peaked around $\alpha$ and $\beta$. Such states can only be
incorporated by our infinite version of the quantum de Finetti
theorem, since they cannot be represented on any tensor product
space $\mathcal{V}^{\otimes n}$, where $\mathcal{V}$ is a
finite-dimensional subspace of $\mathcal{H}$. We note that several
current experiments dealing with systems of many identical particles
\cite{endnote21,Cir06} can produce continuous cat states of the form
$\int d\gamma\, c_\gamma |\gamma\rangle^{\otimes n}$. Some examples
include double-well Bose-Einstein condensates \cite{CLA02},
superconducting current loops \cite{VDW00, Fri00}, and
spin-polarised atomic ensembles \cite{Jul01}.

The Letter is structured as follows. Having defined in more detail
the family of subsets $\{\C_n\}$ to which our theorem applies, we
state our theorem and give an outline of the steps involved in
proving it. The latter parts consider the details of the proof, and
we conclude with some remarks about the extension of this theorem to
larger symmetric subspaces.

\section{Coherent States} We begin by introducing the states which we shall study.
Consider a finite collection of $n$ quantum systems each with a
countably infinite Hilbert space $\H$, such as $n$ modes of the
electromagnetic field. Let $\ket{0}, \ket{1}, \ldots$ label an
arbitrary basis for $\H$. A tensor-product basis state
$\ket{\bm{x}}$ of $\H^{\otimes n}$, labelled by the {\it word}
$\bm{x}$, is given by $\ket{\bm{x}} = \ket{x_1}\otimes \ldots
\otimes \ket{x_n}$.  Let the annihilation operator $a$ be the
operator that lowers the basis states according to $ a \ket{j} =
\sqrt{j}\ket{j-1}$, generalizing the standard definition for
harmonic oscillators. For $\alpha \in \mathbb{C}$, a coherent state
$\ket{\alpha}$ is the eigenstate of the annihilation operator with
eigenvalue $\alpha$ \cite{Louis}.  In this Letter, we prove a de
Finetti theorem for the span \(\C_n\) of {\it coherent power states}
\(\ket{\alpha}^{\otimes{n}}\).

\section{An equivalent description of $\C_n$}
The above definition of \(\C_n\) is in terms of the {\it
overcomplete} spanning set \(\{\ket{\alpha}^{\otimes{n}} |\, \alpha
\in \mathbb{C}\}\), and so does little to elucidate the states that
it encompasses. We shall provide a brief alternative definition in
terms of a basis which is closely linked to the {multinomial} basis
for the symmetric subspace.

The space $\S_n$ is defined to be the span of states $\ket{w}$,
given by\begin{equation}\label{eqn:w}
  \ket{w} = \sqrt{\frac{1}{n^w}}\left(\sum_{\bm{y}, \sum y_i = w} \left.
  {\sqrt{\frac{w!}{y_1!\ldots y_n!}}}\right.
  \ket{\bm{y}}\right),
\end{equation}
for all $w \in \mathbb{Z}, w \geq 0$.  It is possible to show that
\mbox{\(\S_n \equiv \C_n\)} \cite{endnote23}, and so the states
\(\{\ket{w}\}\) provide an orthonormal {\it countable} basis of the
coherent power subspace. They are analogous to the classical urn
model of \(w\) tosses of a fair \(n\)-sided coin.

This second characterization allows us to gain an insight into the
relative size of $C_n$ and the full symmetric subspace, $\Sym
\H^{\otimes n}$. To do so, we define the $w$-weighted subspace to be
the symmetric span of all states $\ket{\bm{x}}$ that satisfy $\sum
x_i = w$ and we note that $\Sym \H^{\otimes n}$ is the direct sum of
these.  The dimension of the $w$-weighted subspaces grows
polynomially with $w$, but in contrast $C_n$ contains only the state
$\ket{w}$ with weight $w$.

\section{A de Finetti theorem} If $\rho_n$ is a pure state on
a particular $\C_n$, we are able to approximate the reduction to $k$
systems $\tr_{n - k}(\rho_n)$ as a mixture of coherent power state
projectors. More precisely, we can construct a probability measure
$\nu(\alpha)$ such that
\begin{equation}\label{eqn:cohdef}
\Delta\equiv\norm{\tr_{n - k}(\rho_n) - \displaystyle\int
\big(\proj\alpha\big)^{\otimes k} \nu(\alpha)\, d^2\alpha}_1 \leq
\frac{3}{2}{\frac{k}{n}}.
\end{equation}
Here, the {\it trace norm} $\norm{\sigma}_1$ is the sum of the
absolute values of the eigenvalues of $\sigma$.

\section{Proof Outline} To prove this theorem, two main
steps are necessary.  We begin by constructing the operator
\begin{equation} \label{eqn:lam_nk}\Lambda_{n,k} = \frac{n - k}{\pi} \int d^2\alpha \,
    I_k \otimes \Big(\proj{\alpha}\Big)^{\otimes n - k},
\end{equation}
where $I_k$ is the identity on the first $k$ subsystems. We show
that its restriction to the span $\C_n$ of coherent power states is
equal to the identity, which transforms the term $\tr_{n - k}
(\rho_n)$, in Eq.~(\ref{eqn:cohdef}), into an integral. This allows
the use of standard inequalities to bound $\Delta$ from above.

The displacement operator, defined below, plays the r\^{o}le of the
unitary transformations from \cite{CHR06}. The crucial fact is that
in our integral $\Lambda_{n,k}$, analogous to that used in
\cite{CHR06}, we have no divergent dimension-dependent factor, which
would be the case for a direct application of the existing proof.
Without this an infinite-dimensional result is not possible.

\section{An identity operator for coherent power states}  We
now wish to show that $\Lambda_{n,k}$ is the identity operator on
$\C_n$. In order to study the multi-mode states, we note that the
\mbox{single-mode} states have an explicit expansion in the basis
$\ket{0}$, $\ket{1}$, $\ldots$.\begin{equation}\label{eqn:cohexp}
  \ket{\alpha} = e^{-\Mod{\alpha}^2/2} \sum_{i=0}^{\infty}
  \frac{\alpha^i}{\sqrt{i!}} \, \ket{i} = D(\alpha) \ket{0},
\end{equation}
where $D(\alpha)$ is the {\textit {displacement operator}},
$D(\alpha) = e^{\alpha a^{\dag} - \bar{\alpha}a}$.

The proof takes two parts. We first show that $\Lambda_{n,k}$ acts
identically upon the vacuum state, and move on to show that it
commutes with the displacement operator. Explicitly we intend to
show{\begin{align}
    \Lambda_{n,k} \ket{\bm{0}} &= \ket{\bm{0}},
    \label{eqn:lamvac}\\
  \left[\Lambda_{n,k},D(\alpha)^{\otimes n}\right] &=0
    \label{eqn:lamalp}\,.
\end{align}}
We prove Eq.~(\ref{eqn:lamvac}) by using the expansion of a coherent
state given in Eq.~(\ref{eqn:cohexp}).  Consider the inner product
of $\Lambda_{n, k}\ket{\bm{0}}$ with a basis vector labelled by a
word $\bm{x}$.
\begin{equation*}
\begin{aligned}&  \bra{\bm{x}}\Lambda_{n, k} \ket{\bm{0}}
  = \\ &  \qquad \frac{n - k}{\pi} \, \delta_{x_1,0}\ldots\delta_{x_k,0}\int d^2\alpha\, e^{-(n - k)\Mod{\alpha}^2}
  \prod_{i=k+1}^n
  \frac{\bar{\alpha}^{x_i}}{\sqrt{x_i!}}.
  \end{aligned}
\end{equation*}
The integral is zero unless $x_i = 0$ for $k < i \leq n$, in which
case we obtain
\begin{equation}
  \bra{\bm{x}} \Lambda_{n, k} \ket{\bm{0}} =
  \begin{cases}
    1 & \text{if }\bm{x} = \bm{0}\\
    0 & \text{otherwise.}
  \end{cases}
\end{equation}
This means that $\Lambda_{n, k}\ket{\bm{0}} = \ket{\bm{0}}$. While
proving Eq.~(\ref{eqn:lamalp}) we employ the notation $D^m_{\alpha}
= D(\alpha)^{\otimes m}$.
\begin{align*}
  &\Lambda_{n, k} \,D_{\alpha}^n = \frac{n -
  k}{\pi}
  \int d^2\beta\, I_k \otimes \left(D_{\beta}^{n-k} \ket{\bm{0}} \bra{\bm{0}}
  D_{-\beta}^{n-k}\right){D_{\alpha}^n}\\
  &\quad = \frac{n-k}{\pi}
  \int d^2\beta \, D_{\alpha}^k  \otimes \left(D_{\beta}^{n-k} \ket{\bm{0}} \bra{\bm{0}}
  \left[D_{-\beta}^{n-k} D_{\alpha}^{n-k}\right]\right)\\
  &\quad =  \frac{n-k}{\pi}\int d^2\gamma \,D_{\alpha}^k \otimes \left(\left[D_{\alpha}^{n-k} D_{\gamma}^{n-k}\right]
  \ket{\bm{0}} \bra{\bm{0}} D_{-\gamma}^{n-k}\right)\\
 &\quad =  \frac{n-k}{\pi}\int d^2\gamma \,D_{\alpha}^n \left(I_k \otimes D_{\gamma}^{n-k}
   \ket{\bm{0}} \bra{\bm{0}} D_{-\gamma}^{n-k}\right)\\
 &\quad = D_{\alpha}^n \Lambda_{n, k}  = D(\alpha)^{\otimes n} \,\Lambda_{n, k},
\end{align*}
where we have substituted $\gamma = \beta - \alpha$ and used the
property {$D(x+y) = e^{\frac{1}{2}(x\bar{y}-\bar x y)} D(x) D(y)$}.

To complete the proof we combine the two previous results,
\begin{equation}\label{eqn:ide}
  \begin{aligned}
    \Lambda_{n, k} \ket{\alpha}^{\otimes n} &= %
    \Lambda_{n, k} \,D(\alpha)^{\otimes n} \ket{\bm{0}}\\
    &= D(\alpha)^{\otimes n} \Lambda_{n, k} \ket{\bm{0}}\\
    &= D(\alpha)^{\otimes n} \ket{\bm{0}} \\
    &= \ket{\alpha}^{\otimes n}.
  \end{aligned}
  \end{equation}

\section{Establishing the bound on $\Delta$}
Having demonstrated that we have a resolution of the identity on
\(\C_n\) we can now begin the proof of the theorem itself. The state
$\rho_n$ is pure, and thus is given by some $\rho_n =
\ket\Psi\bra\Psi$. For each $\alpha$ we define a non-normalized
state on the first $k$ subsystems
\begin{equation*}
\ket{\Psi_k^\alpha} = \sqrt{\frac{n-k}{\pi}}\bigg(I_k \otimes
\bra{\alpha}^{\otimes n - k}\bigg) \ket\Psi,
\end{equation*}
with corresponding positive operator
\begin{equation}
\begin{aligned}
\rho^\alpha_k &= \proj{\Psi_k^\alpha}  \\
&=\frac{n-k}{\pi} \, \tr_{n-k} \bigg(I_k \otimes
\Big(\proj{\alpha}\Big)^{\otimes n - k}\proj{\Psi}\bigg).
\end{aligned}
\end{equation}
Since $\ket{\Psi} \in \C_n$, we note that
    \begin{equation}\label{eqn:normal}
      \int \rho^\alpha_k\, d^2\alpha = \tr_{n-k} \left(\Lambda_{n, k} \proj{\Psi}\right) = \tr_{n-k}({\rho_n}).
    \end{equation}

To define the measure $\nu(\alpha)$ in Eq.~(\ref{eqn:cohdef}), we
further project the states $\rho_k^\alpha$ onto
    $P^\alpha = \proj{\alpha}^{\otimes k}$ and define $\nu(\alpha) = \tr
    (P^\alpha \rho_k^\alpha)$, so that
    $\proj\alpha^{\otimes k} \nu(\alpha) = P^\alpha \rho_k^\alpha
    P^\alpha$.
We then have
\[
\begin{aligned}
  \Delta = &\mathop{\phantom{+}}\norm{\int d^2\alpha\,\left( \rho_k^\alpha - P^\alpha
\rho_k^\alpha P^\alpha\right)}_1\\[2mm]
\leq & \mathop{\phantom{+}} \underbrace{\norm{\int d^2\alpha\,\left(
\rho_k^\alpha - P^\alpha
\rho_k^\alpha\right)}_1}_{\textstyle \zeta} \\& \mathop{+}%
 \underbrace{\norm{\int d^2\alpha\,\left( \rho_k^\alpha
- \rho_k^\alpha P^\alpha\right)}_1}_{\textstyle \eta} \\& \mathop{+} %
\underbrace{\norm{\int d^2\alpha\,(I_k - P^\alpha)\rho_k^\alpha(I_k
- P^\alpha)}_1}_{\textstyle \theta},
\end{aligned} \]
using an identity presented in \cite{CHR06}---
\[
A - BAB = (A - BA) + (A - AB) - (I - B)A(I-B).
\]

And so it is necessary to calculate bounds for \(\zeta\), \(\eta\)
and \(\theta\).  We may do so by employing the completeness relation
that \(\Lambda_{n,k}\) provides.
\[
\begin{aligned}
  \zeta &= \norm{\tr_{n-k}(\rho_n) - \int d^2\alpha\,\frac{n-k}{\pi} %
  \tr_{n-k}\left(\proj{\alpha}^{\otimes n} \rho_n\right)}_1\\
  &= \norm{\tr_{n-k}(\rho_n) - \frac{n-k}{n}%
  \tr_{n-k}\big(\Lambda_{n,0} \rho_n\big)}_1\\
  &= \left(1 - \frac{n-k}{n}\right)\norm{\tr_{n-k}\Big(\rho_n\Big)}_1 = \frac{1}{2}\frac{k}{n}.\\
\end{aligned}
\]

Similarly, we have that \(\eta = \frac{1}{2}\frac{k}{n}\).  Bounding
\(\theta\) is only marginally more complicated.  Beginning with the
triangle inequality,
\[
\begin{aligned}
  \theta &\leq \int d^2\alpha\,\norm{(I_k -
P^\alpha)\rho_n^\alpha(I_k - P^\alpha)}_1\\
&\leq \frac{1}{2}\int d^2\alpha\,\tr\Mod{(I_k -
P^\alpha)\rho_n^\alpha(I_k - P^\alpha)}.
\end{aligned}
\]
Now, since we have the projector, \(I_k - P^\alpha\), straddling a
completely positive operator, \(\rho_n^\alpha\), this simplifies to
\[
\begin{aligned}
 \theta %
 &\leq \frac{1}{2} \int d^2\alpha\,\tr\left((I_k - P^\alpha)\rho_n^\alpha\right)\\
&\leq \frac{1}{2} \tr \left(\int d^2\alpha\,(I_k - P^\alpha)\rho_n^\alpha\right)\\
&\leq \frac{1}{2} \tr \left(\left(1-\frac{n-k}{n}\right)\tr_{n-k}\rho_n\right)=\frac{1}{2} \frac{k}{n}.\\
\end{aligned}
\]
Bringing this all together gives us the final bound since \( \Delta
\leq \zeta + \eta + \theta\).

\section{Conclusion}
We have stated and proved a de Finetti theorem for a limited class
of finitely exchangeable states in a Hilbert space of countably
infinite dimension. The counterexample given in \cite{CHR06} shows
that we shall never have a direct generalisation of the classical
scenario. An important question concerns the characterization of the
set of finitely exchangeable states for which approximate de Finetti
representations do exist. Our work provides a partial answer to this
question for a class of states to which the previously known de
Finetti theorems do not apply.

It is possible to extend our results to a larger class of states by
considering the countably infinite Hilbert space $\H$ as a tensor
product of two, or possibly more, subsystems.  For example, $\H$ is
equivalent to a tensor product of itself and a qubit, providing both
a new family of de Finetti states and a new bound on $\Delta$. This
is the subject of ongoing work.

This work was supported by the UK Engineering and Physical Sciences
Research Council. We thank Renato Renner, Robert K\"onig, and Graeme
Mitchison for useful discussions.


\begin{thebibliography}{20}
\expandafter\ifx\csname
natexlab\endcsname\relax\def\natexlab#1{#1}\fi
\expandafter\ifx\csname bibnamefont\endcsname\relax
  \def\bibnamefont#1{#1}\fi
\expandafter\ifx\csname bibfnamefont\endcsname\relax
  \def\bibfnamefont#1{#1}\fi
\expandafter\ifx\csname citenamefont\endcsname\relax
  \def\citenamefont#1{#1}\fi
\expandafter\ifx\csname url\endcsname\relax
  \def\url#1{\texttt{#1}}\fi
\expandafter\ifx\csname urlprefix\endcsname\relax\def\urlprefix{URL
}\fi \providecommand{\bibinfo}[2]{#2}
\providecommand{\eprint}[2][]{\url{#2}}

\bibitem[{\citenamefont{de~Finetti}(1937)}]{dFi37}
\bibinfo{author}{\bibfnamefont{B.}~\bibnamefont{de~Finetti}},
  \bibinfo{journal}{Ann. Inst. H. Poincar\'{e}} \textbf{\bibinfo{volume}{7}},
  \bibinfo{pages}{1} (\bibinfo{year}{1937}).

\bibitem[{\citenamefont{Diaconis and Freedman}(1980)}]{Dia80}
\bibinfo{author}{\bibfnamefont{P.}~\bibnamefont{Diaconis}} \bibnamefont{and}
  \bibinfo{author}{\bibfnamefont{D.}~\bibnamefont{Freedman}},
  \bibinfo{journal}{Ann. Prob.} \textbf{\bibinfo{volume}{8}},
  \bibinfo{pages}{745} (\bibinfo{year}{1980}).

\bibitem[{\citenamefont{Savage}(1972)}]{Savage1972}
\bibinfo{author}{\bibfnamefont{L.~J.} \bibnamefont{Savage}},
  \emph{\bibinfo{title}{The Foundations of Statistics}}
  (\bibinfo{publisher}{Dover}, \bibinfo{address}{New York},
  \bibinfo{year}{1972}), \bibinfo{edition}{2nd} ed.

\bibitem{endnote}{A power distribution $Y$ of $n$ variables is a
probability density function given by $Y(x_1, x_2,\protect \ldots ,
x_n) = X(x_1)X(x_2)\protect \ldots  X(x_n)$ where $X$ is a
single-variable density function.}

\bibitem{endnote24}{A power state on an $n$-mode system is one that may
be written in the form $\sigma ^{\otimes n}$, where $\sigma$ is a
single-mode density operator.}

\bibitem[{\citenamefont{Hudson and Moody}(1976)}]{HMo76}
\bibinfo{author}{\bibfnamefont{R.~L.} \bibnamefont{Hudson}} \bibnamefont{and}
  \bibinfo{author}{\bibfnamefont{G.~R.} \bibnamefont{Moody}},
  \bibinfo{journal}{Z. Wahrschein. verw. Geb.} \textbf{\bibinfo{volume}{33}},
  \bibinfo{pages}{343} (\bibinfo{year}{1976}).

\bibitem[{\citenamefont{Caves et~al.}(2002)\citenamefont{Caves, Fuchs, and
  Schack}}]{CFS02}
\bibinfo{author}{\bibfnamefont{C.~M.} \bibnamefont{Caves}},
  \bibinfo{author}{\bibfnamefont{C.~A.} \bibnamefont{Fuchs}}, \bibnamefont{and}
  \bibinfo{author}{\bibfnamefont{R.}~\bibnamefont{Schack}},
  \bibinfo{journal}{J. Math. Phys.} \textbf{\bibinfo{volume}{43}},
  \bibinfo{pages}{4537} (\bibinfo{year}{2002}),
  \bibinfo{note}{\href{http://arxiv.org/abs/quant-ph/0104088}{quant-ph/0104088%
}}.

\bibitem[{\citenamefont{St{\o}rmer}(1969)}]{Stormer}
\bibinfo{author}{\bibfnamefont{E.}~\bibnamefont{St{\o}rmer}},
  \bibinfo{journal}{J. Funct. Anal} \textbf{\bibinfo{volume}{3}},
  \bibinfo{pages}{48} (\bibinfo{year}{1969}).

\bibitem[{\citenamefont{Petz}(1990)}]{Petz}
\bibinfo{author}{\bibfnamefont{D.}~\bibnamefont{Petz}}, \bibinfo{journal}{Prob.
  Th. Rel. Fields} \textbf{\bibinfo{volume}{85}}, \bibinfo{pages}{1}
  (\bibinfo{year}{1990}).

\bibitem[{\citenamefont{K{\"{o}}nig and Renner}(2005)}]{Koe05}
\bibinfo{author}{\bibfnamefont{R.}~\bibnamefont{K{\"{o}}nig}} \bibnamefont{and}
  \bibinfo{author}{\bibfnamefont{R.}~\bibnamefont{Renner}},
  \bibinfo{journal}{J. Math. Phys} \textbf{\bibinfo{volume}{46}},
  \bibinfo{pages}{122108} (\bibinfo{year}{2005}),
  \bibinfo{note}{\href{http://arxiv.org/abs/quant-ph/0410229}{quant-ph/0410229%
}}.

\bibitem[{\citenamefont{Brun et~al.}(2001)\citenamefont{Brun, Caves, and
  Schack}}]{BrunEtAl}
\bibinfo{author}{\bibfnamefont{T.~A.} \bibnamefont{Brun}},
  \bibinfo{author}{\bibfnamefont{C.~M.} \bibnamefont{Caves}}, \bibnamefont{and}
  \bibinfo{author}{\bibfnamefont{R.}~\bibnamefont{Schack}},
  \bibinfo{journal}{Phys. Rev. A} \textbf{\bibinfo{volume}{63}},
  \bibinfo{pages}{042309} (\bibinfo{year}{2001}),
  \bibinfo{note}{\href{http://www.arxiv.org/abs/quant-ph/0010038}{quant-ph/001%
0038}}.

\bibitem[{\citenamefont{Christandl et~al.}(2006)\citenamefont{Christandl,
  K{\"{o}}nig, Mitchison, and Renner}}]{CHR06}
\bibinfo{author}{\bibfnamefont{M.}~\bibnamefont{Christandl}},
  \bibinfo{author}{\bibfnamefont{R.}~\bibnamefont{K{\"{o}}nig}},
  \bibinfo{author}{\bibfnamefont{G.}~\bibnamefont{Mitchison}},
  \bibnamefont{and} \bibinfo{author}{\bibfnamefont{R.}~\bibnamefont{Renner}}
  (\bibinfo{year}{2006}),
  \bibinfo{note}{\href{http://arxiv.org/abs/quant-ph/0602130}{quant-ph/0602130%
}}.

\bibitem[{\citenamefont{Schack and Fuchs}(2004)}]{Bayes}
\bibinfo{author}{\bibfnamefont{R.}~\bibnamefont{Schack}} \bibnamefont{and}
  \bibinfo{author}{\bibfnamefont{C.~A.} \bibnamefont{Fuchs}}, in
  \emph{\bibinfo{booktitle}{Quantum State Estimation}}, edited by
  \bibinfo{editor}{\bibfnamefont{M.~G.~A.} \bibnamefont{Paris}}
  \bibnamefont{and}
  \bibinfo{editor}{\bibfnamefont{J.}~\bibnamefont{\v{R}eh\'{a}\v{c}ek}}
  (\bibinfo{publisher}{Springer-Verlag}, \bibinfo{address}{Berlin},
  \bibinfo{year}{2004}), p. \bibinfo{pages}{147},
  \bibinfo{note}{\href{http://www.arxiv.org/abs/quant-ph/0404156}{quant-ph/040%
4156}}.

\bibitem[{\citenamefont{Fuchs et~al.}(2004)\citenamefont{Fuchs, Schack, and
  Scudo}}]{ScudoEtAl}
\bibinfo{author}{\bibfnamefont{C.~A.} \bibnamefont{Fuchs}},
  \bibinfo{author}{\bibfnamefont{R.}~\bibnamefont{Schack}}, \bibnamefont{and}
  \bibinfo{author}{\bibfnamefont{P.~F.} \bibnamefont{Scudo}},
  \bibinfo{journal}{Phys. Rev. A} \textbf{\bibinfo{volume}{69}},
  \bibinfo{pages}{062305} (\bibinfo{year}{2004}),
  \bibinfo{note}{\href{http://www.arxiv.org/abs/quant-ph/0307198}{quant-ph/030%
7198}}.

\bibitem[{\citenamefont{Renner}(2005)}]{Ren05}
\bibinfo{author}{\bibfnamefont{R.}~\bibnamefont{Renner}}, Ph.D. thesis,
  \bibinfo{school}{Zurich} (\bibinfo{year}{2005}),
  \bibinfo{note}{\href{http://arxiv.org/abs/quant-ph/0512258}{quant-ph/0512258%
}}.

\bibitem[{\citenamefont{Renner et~al.}(2005)\citenamefont{Renner, Gisin, and
  Kraus}}]{Gis05}
\bibinfo{author}{\bibfnamefont{R.}~\bibnamefont{Renner}},
  \bibinfo{author}{\bibfnamefont{N.}~\bibnamefont{Gisin}}, \bibnamefont{and}
  \bibinfo{author}{\bibfnamefont{B.}~\bibnamefont{Kraus}},
  \bibinfo{journal}{Phys. Rev. A} \textbf{\bibinfo{volume}{72}},
  \bibinfo{pages}{012332} (\bibinfo{year}{2005}),
  \bibinfo{note}{\href{http://arxiv.org/abs/quant-ph/0502064}{quant-ph/0502064%
}}.

\bibitem[{\citenamefont{Korsbakken et~al.}(2006)\citenamefont{Korsbakken,
  Whaley, DuBois, and Cirac}}]{Cir06}
\bibinfo{author}{\bibfnamefont{J.~I.} \bibnamefont{Korsbakken}},
  \bibinfo{author}{\bibfnamefont{K.~B.} \bibnamefont{Whaley}},
  \bibinfo{author}{\bibfnamefont{J.}~\bibnamefont{DuBois}}, \bibnamefont{and}
  \bibinfo{author}{\bibfnamefont{J.~I.} \bibnamefont{Cirac}}
  (\bibinfo{year}{2006}),
  \bibinfo{note}{\href{http://arxiv.org/abs/quant-ph/0611121}{quant-ph/0611121%
}}.

\bibitem{endnote21}{These systems can be modelled as a collection
of $n$ harmonic oscillator modes.}


\bibitem[{\citenamefont{Claussen et~al.}(2002)\citenamefont{Claussen, Donley,
  Thompson, and Wieman}}]{CLA02}
\bibinfo{author}{\bibfnamefont{N.~R.} \bibnamefont{Claussen}},
  \bibinfo{author}{\bibfnamefont{E.~A.} \bibnamefont{Donley}},
  \bibinfo{author}{\bibfnamefont{S.~T.} \bibnamefont{Thompson}},
  \bibnamefont{and} \bibinfo{author}{\bibfnamefont{C.~E.}
  \bibnamefont{Wieman}}, \bibinfo{journal}{Nature}
  \textbf{\bibinfo{volume}{417}}, \bibinfo{pages}{529} (\bibinfo{year}{2002}).

\bibitem[{\citenamefont{van~der Wal et~al.}(2000)\citenamefont{van~der Wal, ter
  Haar, Wilhelm, Schouten, Harmans, Orlando, Lloyd, and Mooij}}]{VDW00}
\bibinfo{author}{\bibfnamefont{C.~H.} \bibnamefont{van~der Wal}},
  \bibinfo{author}{\bibfnamefont{A.~C.~J.} \bibnamefont{ter Haar}},
  \bibinfo{author}{\bibfnamefont{F.~K.} \bibnamefont{Wilhelm}},
  \bibinfo{author}{\bibfnamefont{R.~N.} \bibnamefont{Schouten}},
  \bibinfo{author}{\bibfnamefont{C.~J. P.~M.} \bibnamefont{Harmans}},
  \bibinfo{author}{\bibfnamefont{T.~P.} \bibnamefont{Orlando}},
  \bibinfo{author}{\bibfnamefont{S.}~\bibnamefont{Lloyd}}, \bibnamefont{and}
  \bibinfo{author}{\bibfnamefont{J.~E.} \bibnamefont{Mooij}},
  \bibinfo{journal}{Science} \textbf{\bibinfo{volume}{290}},
  \bibinfo{pages}{773} (\bibinfo{year}{2000}).

\bibitem[{\citenamefont{Friedman et~al.}(2000)\citenamefont{Friedman, Patel,
  Chen, Tolpygo, and Lukens}}]{Fri00}
\bibinfo{author}{\bibfnamefont{J.~R.} \bibnamefont{Friedman}},
  \bibinfo{author}{\bibfnamefont{V.}~\bibnamefont{Patel}},
  \bibinfo{author}{\bibfnamefont{W.}~\bibnamefont{Chen}},
  \bibinfo{author}{\bibfnamefont{S.~K.} \bibnamefont{Tolpygo}},
  \bibnamefont{and} \bibinfo{author}{\bibfnamefont{J.~E.}
  \bibnamefont{Lukens}}, \bibinfo{journal}{Nature}
  \textbf{\bibinfo{volume}{406}}, \bibinfo{pages}{43} (\bibinfo{year}{2000}).

\bibitem[{\citenamefont{Julsgaard et~al.}(2001)\citenamefont{Julsgaard,
  Kozhekin, and Polzik}}]{Jul01}
\bibinfo{author}{\bibfnamefont{B.}~\bibnamefont{Julsgaard}},
  \bibinfo{author}{\bibfnamefont{A.}~\bibnamefont{Kozhekin}}, \bibnamefont{and}
  \bibinfo{author}{\bibfnamefont{E.~S.} \bibnamefont{Polzik}},
  \bibinfo{journal}{Nature} \textbf{\bibinfo{volume}{413}},
  \bibinfo{pages}{400} (\bibinfo{year}{2001}).

\bibitem[{\citenamefont{Louisell}(1990)}]{Louis}
\bibinfo{author}{\bibfnamefont{W.~H.} \bibnamefont{Louisell}},
  \emph{\bibinfo{title}{Quantum Statistical Properties of Radiation}}
  (\bibinfo{publisher}{Wiley}, \bibinfo{address}{New York},
  \bibinfo{year}{1990}).

\bibitem{endnote23}{To prove this we use the projector $\Lambda _{n,0}$
defined in Eq.~(\ref{eqn:lam_nk}). This allows us to write
$\ket{w}~=~\int \,d^2\beta\,w_\beta \ket{\beta}^{\otimes n}$ and
$\ket{\alpha}^{{\otimes {n}}} = \sum \alpha _{w'} \ket{w'}$ for all
$\alpha $ and $w$, giving $\mathcal {S}_n \subset \mathcal {C}_n
\subset \mathcal {S}_n$.}

\end{thebibliography}
\end{document}